\documentstyle[preprint,tighten,aps,psfig]{revtex} 
\begin{document}           %
\draft
\preprint{\vbox{\noindent
To appear in Phys. Rev. D\null \hfill astro-ph/9902222\\
          \null\hfill  INFNCA-TH9814}}
\title{Solar neutrino fluxes with arbitrary $^3$He mixing \\
      }
\author{ 
         V.~Berezinsky,$^{1,}$\cite{email1},
         G.~Fiorentini$,^{2,}$\cite{email2},
         and M.~Lissia$^{3,}$\cite{email3}
       }
\address{
$^{1}$ Istituto Nazionale di Fisica Nucleare, Laboratori Nazionali del
       Gran Sasso, \\ I-67010 Assergi (AQ), Italy \\ 
$^{2}$ Dipartimento di Fisica dell'Universit\`a di Ferrara and \\
        Istituto Nazionale di Fisica Nucleare, 
        Sezione di Ferrara, I-44100 Ferrara, Italy \\
$^{3}$ Istituto Nazionale di Fisica Nucleare, Sezione di Cagliari and \\
        Dipartimento di Fisica dell'Universit\`a di Cagliari,
        I-09042 Monserrato (CA), Italy
        }
\date{January 24, 1999; revised August 10, 1999}
\maketitle                 
\begin{abstract}
The $^3$He abundance is not constrained by helioseismic data.
Mixing of $^3$He inside the solar core by processes not included in the 
solar standard model (SSM)
has been recently proposed as possible solution to the solar neutrino problem.
We have performed a model independent analysis of solar 
neutrino fluxes using practically arbitrary $^3$He mixing. In addition,
we have been simultaneously varying within very wide ranges the 
temperature in the neutrino production zone and the astrophysical factors,
$S_{17}$ and $S_{34}$, of the $p+{}^7$Be and
$^3$He + $^4$He cross sections.
Seismic data are used as constraints, but the solar-luminosity constraint 
is not imposed. It is demonstrated that even allowing $^3$He abundances
higher by factors up to 16 than in the SSM, temperatures 5\% (or more) lower,
the astrophysical factor $S_{17}$ up to 40\% higher, and varying $S_{34}$
in the range (-20\% ,+40\%), the best fit is still more than
$5 \sigma$ away from the observed fluxes. We conclude that practically 
arbitrary $^3$He mixing combined with independent variations of
temperature and cross-sections cannot explain the observed solar neutrino 
fluxes.
\end{abstract}
%
%
\narrowtext
\section{Introduction}
\label{intro}
Detected neutrino fluxes in all five solar neutrino experiments 
(Homestake~\cite{REF_Hom}, SAGE~\cite{REF_Sa}, GALLEX~\cite{REF_Ki},
Kamiokande~\cite{REF_Kam}, and Superkamiokande~\cite{REF_SKAM})
are smaller than those predicted by SSM's. The status of
the solar neutrino measurements is that of {\em disappearance oscillation
experiments}. 

How reliable is this conclusion? In other words, how reliable are the
SSM predictions of solar neutrino fluxes?

1. {\em Helioseismic observations} of density and sound-speed-squared profiles 
are in agreement with SSM predictions throughout the sun with an accuracy
better than a fraction of percent~\cite{REF_CD}.
This agreement was recently found to be valid also in the inner solar core, 
$R \leq 0.1 R_{\odot}$, where neutrino fluxes are mostly produced 
\cite{REF_Dz,REF_Fi,REF_Basu,BBP98,BTCM98}. 

In the literature the relevance of this agreement for the neutrino fluxes
has been questioned with two kinds of objections.

First, seismic measurements give only the sound speed squared, $c_s^2$, but
not directly the temperature $T$, which is mostly relevant for neutrino
production. The connection between the uncertainties of $c_s^2$ and
$T$ is $\delta c_s^2/c_s^2 = \delta T /T- \delta\mu/\mu$, where $\mu$ and
$\delta\mu$ are the molecular weight and its uncertainty.
In principle, one can imagine that $\delta T/T$ much larger
than $\delta c_s^2/c_s^2$ could be compensated by correlated changes of
the molecular weight. Note, however,
that this compensation (fine tuning) should work for all distances, and
this is harder to imagine. 

Second, seismic uncertainties grow for $R < 0.05 R_{\odot}$. 
Nevertheless, these uncertainties are too small to solve
solar-neutrino problem. Careful analyses of the inversion method
and of the available helioseismic data produce $c_s^2$ in the agreement  
with the SSM value at level of $1\%$ at $R=0$, and better in the neutrino
production zone~\cite{Fi}.

2. {\em Nuclear cross-sections} for neutrino production could be 
responsible for the observed neutrino deficit. 

Recent measurements~\cite{LUNA} of the $^3$He + $^3$He cross section at
the energy of the Gamow peak in the sun have eliminated a dangerous source of 
potential uncertainty. Moreover, even before this experiment, it had
been already demonstrated that any combination of unknown cross sections
and variations of central temperature
cannot not explain the combined results of solar-neutrino
experiments~\cite{BeFiLi}.

As everything in the world, SSM is not perfect. It does not include, 
for example, rotation and hydrodynamical processes. A bump in the
sound-speed profile at $R \approx 0.7 R_{\odot}$ shows
statistically significant disagreement between SSM and helioseismic
observations. Nevertheless, this disagreement is far too small to 
affect neutrino fluxes. It is also most likely that rotation and
hydrodynamical processes do not affect neutrino fluxes either. Indeed,
in the region where statistically significant deviations from the SSM
predictions are seen, helioseismology is a much more sensitive probe
of the solar structure than neutrino experiments.

Lithium depletion is another problem for SSM's. The observed surface 
abundance of lithium in the sun (and other main sequence stars) is much
lower than the one predicted by SSM's. This problem can be solved by anomalous 
diffusion, which drives lithium inside the sun, where it burns. This
diffusion could be induced by gravity waves~\cite{Sch}.

Gravity-wave induced diffusion is a mechanism of mixing inside the 
solar core. Another mixing mechanism, periodical instabilities, has been
considered again and again in the literature starting 
with the pioneering work by Dilke and Gough~\cite{DG}.
Mixing in the solar core affects the solar neutrino fluxes, but it should be
shown to be consistent with the other data; in general, large mixing
contradicts seismic observations. Indeed, mixing episodes transport
from the periphery to the center of the sun not only the desirable 
$^3$He or lithium but also large masses of hydrogen.
The consequent reduction of the molecular weight increases the sound speed.
This is the main problem of the Cummings-Haxton assumption~\cite{CummHaxt96} 
about
mixing in the solar core: the resulting sound speed is too high~\cite{BPBCD}.
Brun {\em et al.}~\cite{BTCM98} recently found that
seismic data do not support even small mixing in the solar core. Using
the turbulent diffusion coefficient suggested by Morel and
Schatzman~\cite{MoSch}, they obtained a sound-speed profile in contradiction
with observations.

The essence of the recent attempts to reconcile the observed neutrino 
fluxes with solar models~\cite{Gough,CummHaxt96,Sch,Haxton} consists in
bringing $^3$He in the solar core by one mechanism or another.
Arguments against these mechanisms use the {\em accompanying} processes,
such as the simultaneous bringing-in of hydrogen. The strategy of our work is 
different: we allow arbitrary mixing of $^3$He, not accompanied
by other elements. Moreover, we do not assume that the arriving $^3$He is
in nuclear equilibrium with the other elements; an example can be
the out-of-equilibrium concentration of $^3$He in mixing episodes
caused by instability. Nonequilibrium burning of $^3$He lowers 
the temperature. We shall allow {\em independent} variation of temperature 
within $\pm 5\%$ of its SSM value, assuming that some miraculous fine-tuning
keeps the sound speed in agreement with seismic observations. In the
solar core we also allow radial profiles of temperature and $^3$He density
different from the SSM ones. In addition, we vary the two relevant
cross-sections, $S_{17}$ and $S_{34}$, within very generous ranges. 

We shall demonstrate that the resulting predictions
are, nonetheless, incompatible with the observed neutrino fluxes.     

\section{Assumptions, constraints and notation}
We shall study the production of boron ($^8$B) and beryllium ($^7$Be)
neutrinos. Our approach consists in comparing the calculated fluxes
($\Phi_B,\Phi_{Be}$) with the observed ones.

Inspired by the SSM, we consider an effective Neutrino Production Zone (NPZ)
for boron and beryllium neutrinos. In the SSM the production of 
B and Be neutrinos has its maximum at  $r=R_{NPZ} \approx 0.05 R_{\odot}$.
When non-SSM profiles are considered, the position of this maximum changes.
In our calculations we use $R_{NPZ}$ only as a reference radius and 
take into account 
that the peak of the B-neutrino production does not coincide with that 
for Be-neutrinos.
 All quantities evaluated at $r=R_{NPZ}$ are denoted by the index $NPZ$,
{\em e.g.}, $T_{NPZ}$ (temperature), $\rho_{NPZ}$ (density),
$Y_{3,NPZ}$ ($^3$He abundance), etc. Temperature, density and abundance
profiles will be specified.

We assume that $^3$He is transported to NPZ by some unspecified process and 
characterize it by the arbitrary value $Y_{3,NPZ}$. We do not assume local 
nuclear equilibrium for $^3$He: during mixing episodes, ``fresh''
$^3$He can be brought inside from the outer shells and burned in a
nonequilibrium regime. However, it is easy to see that thermal equilibrium is 
established very fast and thus $^3$He has the thermal distribution of
the surrounding gas.  

We keep $Y_{3,NPZ}$ and $T_{NPZ}$ as independent parameters.  
In a realistic solar model increasing $Y_{3,NPZ}$ increases the
$^3$He + $^3$He reaction rate more than the rate of the $^3$He + $^4$He
reaction, and the temperature must be lowered to keep the luminosity
constant; the net result would be the decreasing of $\Phi_{Be}$ and
$\Phi_{B}$. 
In fact, in the SSM one finds that $Y_{3}\sim T^{-7}$. For our purpose,
any dependence between $Y_{3,NPZ}$ and $T_{NPZ}$ imposes additional 
restrictions and makes  our conclusions stronger.
 
We do not impose the luminosity constraint. The luminosity
sum rule is only used to express the $pp$ neutrino contribution to the gallium
experiments in terms of the other fluxes. 

The CNO neutrino flux is conservatively neglected to avoid  
solar-model-dependent consideration. Including the CNO neutrino production 
would only strengthen our conclusions.

{\em Helioseismic constraints} are used very conservatively. We assume 
that temperature in the solar core can differ from that of the SSM
by $\pm 5\%$, to be compared with a fraction of percent allowed for
the sound speed squared $c_s^2$.
More specifically, we assume 
the following maximum range of temperature variation:
 \begin{equation}
     0.95  < T_{NPZ}/T^{SSM}_{NPZ} < 1.05 \, ,
\label{temp}
\end{equation}
and the same order of variation for the radial profile of the temperature
in the NPZ.

The sound speed squared is given by $ c_s^2 = \gamma R T / \mu $, where
$\gamma$ and $R$ are constant, $T$ is the temperature, and $\mu$ the
mean molecular weight, $1/\mu = 2 X + 3 Y / 4 + Z / 2 
\approx (5/4) (X + 3/5)$. Therefore, since
the accuracy of seismic determination of the sound speed is
better than 1\%, the following constraint must be met:
\begin{equation}
\frac{X(r) + 3/5}{X^{SSM}(r) + 3/5} = \frac{T^{SSM}(r)}{T(r)} \, .
\end{equation}
Since the allowed variation of $T(r)$ is only a few percent, one can use
\begin{equation}
X(r) = X^{SSM}(r)
\end{equation}
with the accuracy needed: a few percent change in the temperature strongly
affects  neutrino production, while the same variation of $X$ does 
not do much.

For the $^4$He abundance we use $Y \approx 1-X$.

We can use the SSM density profile, $\rho(r)$, since this profile has been
measured by seismic observations.

{\em The astrophysical factors} that affect B and Be neutrino production are 
$S_{17}$ and $S_{34}$.

For the reaction $p+{}^7$Be the INT Collaboration~\cite{INT} (1998) suggests
the $1\sigma$ range $S_{17} =19^{+4}_{-2}$~eV~b;
Castellani {\em et al.}~\cite{report97} (1997)
suggest $S_{17} = 22.4 \pm 2.1$~eV~b.
The BBP98 SSM~\cite{BBP98} and the 
BTCM SSM~\cite{BTCM98} use $19$~eV~b, while the DS96
SSM~\cite{DS96} uses $17$~eV~b.
We shall use the generous range: 

\begin{mathletters}
\label{limit}
\begin{equation}
13<S_{17}<31 \text{ eV b}.
\end{equation}

For the reaction ${}^3$He + ${}^4$He the INT Collaboration~\cite{INT}
suggests the $1\sigma$ range $S_{34} =0.53\pm 0.05$~keV~b; Castellani
{\em et al.}~\cite{report97} suggest $S_{34} = 0.48 \pm 0.02$~keV~b.
The BBP98 SSM~\cite{BBP98} and the BTCM SSM~\cite{BTCM98} use $0.53$~keV~b,
while the DS96 SSM~\cite{DS96} uses $0.45$~keV~b.
We consider the range: 
\begin{equation}
 0.38 < S_{34} < 0.68 \text{ keV b}.
\end{equation}
\end{mathletters}

The {\em detected neutrino fluxes} used in our analysis are
the following.

For the Chlorine signal we use the Homestake data
$2.56 \pm 0.21$~SNU~\cite{REF_Hom};
for B-neutrino flux the Superkamiokande data
$\Phi_{B} = (2.46 \pm 0.09)\times 10^6$
cm$^{-2}$s$^{-1}$~\cite{REF_SKAM}; and 
for gallium signal we use the weighted average of the
GALLEX~\cite{REF_Ki} and 
SAGE~\cite{REF_Sa} data $72.5 \pm 5.7$~SNU.

\section{Parameterization of neutrino production rates}
                                       
The rate of Be neutrino production as function of the radius is:
\begin{mathletters}
\begin{eqnarray}
\label{qbera}
Q_{\nu_{Be}} (r) &=&  4 \pi r^2 \, n_e(r)  \, n_7(r) \, \lambda_{e7}(T(r)) \\
    &=& 4 \pi r^2 \, P_{Be}(r) \, n_3(r) \, n_4(r) \, \lambda_{34}(T(r)) \, ,
\label{qberb}
\end{eqnarray}
where $n_x$ ($x=1,3,4,7,e$) is the number density of particles 
(protons, ${}^3$He, ${}^4$He, ${}^7$Be nuclei, electrons),
$P_{Be} =\lambda_{e7}\, n_e/[\lambda_{e7}\, n_e + \lambda_{17}\, n_1] $ 
is the electron capture probability and
$\lambda_{ij} = \langle v \sigma_{ij} \rangle$ is the reaction rate 
averaged over the Maxwellian distribution of the relative
velocities, which is a know function of temperature and does not depend 
on solar model.

Since the $^7$Be electron-capture-lifetime in the solar core is very short,
$\tau_e({}^7Be) \approx 1$ yr, 
the equilibrium value 
$n_7=n_3n_4\lambda_{34}/(n_e \lambda_{e7}+n_1\lambda_{17})$ 
has been substituted into Eq.~(\ref{qbera})
to obtain Eq.~(\ref{qberb}). With $ P_{Be} \approx 1 $,
the rate of Be neutrino production becomes:
\begin{equation}
\label{qberc}
Q_{\nu_{Be}} (r) =  \frac{\pi N_A^2}{3}\, (1-X(r)) \,
          \left[r \rho(r)\right]^2 
          \, \lambda_{34}(T(r) )\, Y_3(r) \, .
\end{equation}
\end{mathletters}

After similar transformations, the rate of B neutrino production,
\begin{mathletters}
\begin{equation}
Q_{\nu_{B}} (r) = 4 \pi r^2 \, n_1(r) \, n_7(r) \, \lambda_{17}(T(r))
= Q_{\nu_{Be}} (r)
   \frac{n_1(r)\,\lambda_{17}(T(r))}{n_e(r) \, \lambda_{e7}(T(r))},
\end{equation}
becomes
\begin{equation}
\label{qbrc}
Q_{\nu_{B}} (r) =  \frac{\pi N_A^2}{3} \frac{2 (1-X(r)) X(r)}{1+X(r)}  \,
            \left[r \rho(r)\right]^2  \,       
          \frac{\lambda_{34}(T(r)) \lambda_{17}(T(r))}{\lambda_{e7}(T(r))}
          \, Y_3(r) \, .
\end{equation}
\end{mathletters}

Let us introduce the following scaling variables:
\begin{mathletters}
\begin{eqnarray}
x &\equiv& r / R_{NPZ} \\
    \tilde{\rho}(x) & \equiv & \rho(r)/ \, \rho_{NPZ} \,   \\
    y_4(x) & \equiv &  \left(1-X(r)\right) /\, \left(1-X_{NPZ}\right) \,  \\
    y_3(x) & \equiv & Y_3(r)/\, Y_{3,NPZ}\,  \\
    \Lambda_{34}(x) & \equiv &  \lambda_{34}(T(r))/\,\lambda_{34}(T_{NPZ})\, .
\end{eqnarray}
\end{mathletters}
The reference radius $R_{NPZ}$ is taken within the Neutrino Production
Zone (NPZ); in practice, we use $R_{NPZ}= 0.0535 R_{\odot}$.

The profile function normalizations are
$\tilde{\rho}(1)= y_4(1)=  y_3(1) = \Lambda_{34}(1) = 1$.
The total rate of Be neutrino production becomes:
\begin{equation}
Q_{\nu_{Be}} =  \frac{\pi N_A^2}{3} \, 
          R_{NPZ}^3 
          \, \rho_{NPZ}^2 
          (1-X_{NPZ}) \, \frac{S_{34}}{S_{34}^{SSM}} \,
          \lambda_{34}(T_{NPZ}) \, Y_{3,NPZ} \, I_{Be} \, ,
\end{equation}
where we have defined the dimensionless integral over the profile functions
\begin{equation}
\label{intprobe}
I_{Be} =  \int_0^{\infty} dx \left[x\rho(x)\right]^2 \, y_4(x) \,
          \Lambda_{34}(x) \, y_3(x) \, ,
\end{equation}
and $\lambda_{34}$ is calculated with the SSM temperature profile
(the dependence on the cross section has been explicitly 
taken into account by $S_{34}$).

In practice, we can use $\rho_{NPZ} = \rho_{NPZ}^{SSM}$ and
$X_{NPZ} = X_{NPZ}^{SSM}$, and in the integral
$\tilde{\rho}(x) = \tilde{\rho}^{SSM}(x)$ and $y_4(x) \approx y_4^{SSM}(x)$,
since these quantities are strongly constrained by helioseismology and the
dependence of the results on their precise value is not strong. 
On the contrary, $\Lambda_{34}(x)$ and $y_3(x)$ are not directly constrained
by helioseismology and can be different from the SSM: therefore, we shall
study the effects of their change.
Therefore, the Be neutrino production rate can be expressed as 
\begin{mathletters}
\label{bebexact}
\begin{equation}
\label{beexact}
\frac{Q_{\nu_{Be}}}{Q^{SSM}_{\nu_{Be}}} =
  \frac{S_{34}}{S^{SSM}_{34}} \,
  \frac{\lambda_{34}(T_{NPZ})}{\lambda_{34}(T^{SSM}_{NPZ})} \, 
  \frac{Y_{3,NPZ}}{Y_{3,NPZ}^{SSM}} \, 
  \frac{I_{Be}}{I^{SSM}_{Be}} \, .
\end{equation}
 Notice that $\lambda_{34}(T)$
is a known function of $T$, independent of the solar model.

Similarly, the B neutrino production rate can be expressed as
\begin{equation}
\label{bexact}
\frac{Q_{\nu_{B}}}{Q^{SSM}_{\nu_{B}}} =
  \frac{S_{34}}{S^{SSM}_{34}} \, 
  \frac{S_{17}}{S^{SSM}_{17}} \, 
  \frac{\eta(T_{NPZ})}{\eta(T^{SSM}_{NPZ})} \, 
  \frac{Y_{3,NPZ}}{Y_{3,NPZ}^{SSM}} \, 
  \frac{I_{B}}{I^{SSM}_{B}} \\
\end{equation}
\end{mathletters}
where
\begin{equation}
\eta(T) \equiv \frac{\lambda_{34}(T)\lambda_{17}(T)}{\lambda_{e7}(T)}
\end{equation}
is another known solar-model-independent function of $T$,
and
\begin{equation}
\label{intprob}
I_{B} =  \int_0^{\infty} dx \left[x\rho(x)\right]^2 \, 
          \frac{y_4(x) \, (1-y_4(x))}{2-y_4(x)} \,
          H(x) \, y_3(x) \, 
\end{equation}
where $H(x)$ is another profile function:
\begin{equation}
\eta(T(r))  \equiv  \eta(T_{NPZ}) H(x) \, ,
\end{equation}
whose normalization is also $H(1)=1$.

At this point, we should discuss the role of the normalization radius
$R_{NPZ}$. Equations~(\ref{bebexact}) show that the only
dependence on $R_{NPZ}$ left is in the profile functions $I$.
As long as we compare models with the same profiles, the functions $I$
cancel and there remains no dependence on $R_{NPZ}$.
However, we are interested also in models with non-SSM profiles. In this 
case variations of the reference radius modify, in principle,  
the ratios $I/I^{SSM}$ and consequently the fluxes, when all
other parameters stay fixed.
The important point is that these changes of fluxes and ratios  can be 
absorbed by appropriate variations of other parameters, in particular of
$T_{NPZ}$ and $Y_{3,NPZ}$, and therefore, a reference radius does not
appear as another free parameter.
In particular, we use such a parameterization of the profile deformation (see
below) that a variation of the reference radius is 
equivalent to an overall multiplicative factor for the quantity whose
profile deformation is considered.

It is also clear that the choice of the SSM is not essential:
Eqs.~(\ref{bebexact}) are valid for any two solar models that satisfy
the helioseismic constraints.

\subsubsection{SSM radial profiles}
We start our analysis by assuming that the shapes of all radial profiles, 
including $T(r)$ and $Y_3(r)$, are those of the SSM. Since the SSM gives 
a good description of the solar structure and since we rescale all 
values at $R_{NPZ}$, this very reasonable assumption should already be
a good approximation.
In this case $I_{Be}=I_{Be}^{SSM}$ and $I_B=I_B^{SSM}$ 
in Eqs.~(\ref{bebexact}). Then one obtains

\begin{mathletters}
\label{parused}
\begin{eqnarray}
\frac{Q_{\nu_{Be}}}{Q^{SSM}_{\nu_{Be}}} &=&
  \frac{S_{34}}{S^{SSM}_{34}} \times
  \left( \frac{T_{NPZ}}{T_{NPZ}^{SSM}} \right)^{17} \times
  \frac{Y_{3,NPZ}}{Y_{3,NPZ}^{SSM}} \\
\frac{Q_{\nu_{B}}}{Q^{SSM}_{\nu_{B}}} &=&
  \frac{S_{34}}{S^{SSM}_{34}} \times
  \frac{S_{17}}{S^{SSM}_{17}} \times
  \left( \frac{T_{NPZ}}{T_{NPZ}^{SSM}} \right)^{30} \times
  \frac{Y_{3,NPZ}}{Y_{3,NPZ}^{SSM}} \, ,
\end{eqnarray}
\end{mathletters}
where, for convenience and without loss of generality, the (model-independent) 
temperature dependence of the two functions $\lambda_{34}(T)$ and 
$\eta(T)$ have been approximated by power-law functions (this approximation
is quite good in the region of interest $T_{NPZ}/T_{NPZ}^{SSM} =1 \pm 0.05$):
$\lambda_{34}(T) \sim T^{17}$ and $\eta (T) \sim T^{30}$. We shall use these
power-law functions from now on.

At this point we are left with four free parameters $S_{34}, S_{17}, T_{NPZ}$ 
and $Y_{3,NPZ}$. Eqs.~(\ref{parused}) show
that changing $Y_{3,NPZ}$ by an overall factor has the same effect
as changing $S_{34}$.
Since $\nu_B$ and $\nu_{Be}$ fluxes are proportional to $Y_{3,NPZ}$,
increasing $Y_{3,NPZ}$ results in the simultaneous increase of both fluxes 
without changing 
the $\Phi_{Be}/\Phi_{B}$ ratio. It does not help to solve SNP,
making actually the problem even harder. 

In Figure~\ref{fig1} the two thin solid curves on both sides of the curve
``temp.'' (thick solid curve) bound the allowed region. This region
is given by Eqs.~(\ref{parused}) when 
$S_{17}$ and $S_{34}$ are varied within the limits of Eqs.~(\ref{limit})
and the temperature is arbitrary. In this region the best fit to the
experimental data is indicated by the small cross labeled 8
($\Phi_{Be}=2.02\times10^9$ cm$^{-2}$s$^{-1}$,
$\Phi_{B}=2.34\times 10^6$ cm$^{-2}$s$^{-1}$) and it is obtained with
$S_{34} = 0.38$~keV~b, $S_{17} = 31$~eV~b and $T_{NPZ}/T_{NPZ}^{SSM} = 0.969$.
Even if we are assuming no CNO flux, its $\chi^2=35$ is still quite large:
more than $5\sigma$ away from the experimentally allowed region.
It is easy to understand from Fig.~\ref{fig1} that further lowering the
temperature below $T_{NPZ}/T_{NPZ}^{SSM} = 0.969$ does not help, unless one
also allows values of $S_{17}$ larger than 31~eV~b (larger upward shifts). 

It is instructive to follow some selected trajectories within the allowed
region.

We can start for example from the BBP98 SSM, {\em i.e.},
the diamond in Figure~\ref{fig1}. 

When the temperature is lowered and the other parameters are kept constant,
the fluxes follow the thick solid line labeled
``temp.''~\footnote{This is not the usual temperature solution, where
the other parameters, and in particular the $^3$He abundance, also change as
function of temperature.}. The heavy dots labeled
1 and 4 correspond to temperatures scaled by 0.98 and 0.95. 
From these points 1 and 4
one can change $S_{34}$ (or equivalently $Y_{3,SSM}$) and reach points 
2 and 6, respectively. Then one reaches points 3 and 7 by
increasing $S_{17}$. Trajectories are shown by dashed lines. 
Going from point 1 to point 2 corresponds to changing $S_{34}$ from
0.53 to 0.38~keV~b, and going from point 2 
to point 3 to changing $S_{17}$ from 19 to 23~eV~b.
In fact, this set of parameters is the best solution, if one keeps
$S_{17}$ within the range of Eqs~(\ref{limit}) and the temperature within
2\% of the SSM value. Its $\chi^2 = 48$: it excludes the experimental
data at more than $6\sigma$. 
For reference, had we taken $\Phi_{CNO}= \Phi_{CNO}^{SSM}$, instead of
$\Phi_{CNO}=0$, the $\chi^2 $ would have been 82, which excludes
the experimental data at more than $8\sigma$.

Two other possible trajectories corresponding to the lower tem\-pe\-ra\-tu\-re
($T_{NPZ}/T_{NPZ}^{SSM} = 0.95$) are the one leading to point 5
($\chi^2 = 80$), which is reached by increasing $S_{17}$ from 19 to its
maximum value 31~eV~b, and the one leading to point 7 ($\chi^2 = 187$),
which is reached by decreasing $S_{34}$ from 0.53 to its minimum value
0.38~keV~b ($4\to 6$) and then increasing $S_{17}$ from 19 to 31~eV~b.
Note, that we described these trajectories only for illustration; numerical 
calculations used directly the parameterizations of Eqs.~(\ref{parused}).

\subsubsection{The modified radial profiles}
Now we generalize the above analysis and allow the shape of the radial 
profiles to be different from that of the SSM. Therefore, we do not
assume anymore
$I_{Be,B}=I_{Be,B}^{SSM}$ and obtain:

\begin{equation}
\label{ratiobeb}
\frac{Q_{\nu_{Be}} / Q^{SSM}_{\nu_{Be}}} 
     {Q_{\nu_{B}} / Q^{SSM}_{\nu_{B}}}
     = 
     \frac{S^{SSM}_{17}}{S_{17}} \times
     \left( \frac{T_{NPZ}}{T_{NPZ}^{SSM}} \right)^{-13} \times
     \frac{I_{Be}/I_{Be}^{SSM}}{I_{B}/I_{B}^{SSM}} \, .
\end{equation}

In comparison with the previous case the only remaining hope is that the 
ratios $I_{Be,B}/I_{Be,B}^{SSM}$, determined by the radial profiles, 
could improve the agreement with the experimental data.

In particular, a large deformation of the ${}^3$He profile could,
in principle, significantly change the integrals $I_{Be,B}$ circumventing
the problem of the too large ratio of boron to beryllium flux; this
approach corresponds to the proposal of Cumming and
Haxton~\cite{CummHaxt96}.

In fact, the only profile functions which can significantly
affect the integrals $I_{Be}$ and $I_{B}$ are
those of the temperature and the $^3$He abundance: the former
because of the strong temperature dependence of the rates, the latter
because of its being basically unconstrained by helioseismology.

To estimate the dependence of these integrals on the temperature and
$^3$He profiles
we introduce the following parameterization:
\begin{equation}
\label{profvar}
 F_{\delta}(r/R_{NPZ}) \equiv F_{SSM}(r/R_{NPZ}) \times 
    \left(\frac{r + 0.0535R_{\odot}}{R_{NPZ}+0.0535R_{\odot}}
     \right)^{\pm\delta} \, ,
\end{equation}
where $F$ is either the temperature or the $^3$He profile.
This parameterization keeps the normalization $F(1)=1$ fixed,
gives larger deformations for larger $|\delta|$'s and reproduces
the SSM profile when $\delta = 0$. Positive (negative)
$\delta$'s increase (decrease) the function for $r > 0.0535 R_{\odot}$
relative to $r < 0.0535 R_{\odot}$. Since we want a single parameter
function, the scale $0.0535 R_{\odot}$ is kept fixed: $\delta$ is sufficient
to control the first derivative of the profile at $R_{NPZ}$. In other
words, this parameterization includes all possible changes of the
overall normalization and first derivative of the temperature and
$^3$He profile.
Parameterizing differently the deformation in the NPZ (more parameters
and/or different functional forms) must lead to similar conclusions,
since the beryllium and boron neutrino production region is sufficiently
small, so that higher order derivatives result  only in small corrections.

In fact, we have explicitly checked that, if we drastically change this
scale from $0.05 R_{\odot}$ to $0.025 R_{\odot}$ (changes in the other
direction are not relevant to the solution of the solar neutrino
problem), this new
parameterization reproduces the same results within a few percent,
obviously for different values of $\delta$'s (for instance, the case
$\delta = -4$ is reproduced by $\delta = -2.7$). The dotted curve in
Fig.~\ref{fig2} shows that the difference between the two parameterizations
is small in the region where neutrinos are produced.
In this respect, we find instructive, and less dependent
on the specific parameterization, to consider the changes of the profile at
two representative points $r=0.01 R_{\odot}$ and $r=0.1 R_{\odot}$ of the
inner and outer part of the NPZ instead of $\delta$ itself.

Notice that changing the reference radius $R_{NPZ}\to R_{NPZ}'$ in
Eq.~(\ref{profvar}) only give an overall multiplicative factor that
can be absorbed in the scale factors $T_{NPZ}$ and $Y_{3,NPZ}$, confirming
that the choice of $R_{NPZ}$ is irrelevant also in the case of deformed
profiles.

{\em Temperature profile} with $|\delta| = 0.057$ in Eq.~(\ref{profvar}) 
results 
in a maximal temperature correction of about 5\% in the relevant region 
($0.01 R_{\odot}< r < 0.1 R_{\odot}$).
In Table~\ref{table1} we report the dependence of 
$I_{Be}$ and $I_{B}$ on the change of shape of the temperature profile. 
It is particularly important the relative change of $I_{Be}/I_{B}$.
The ratio $I_{Be}/I_{B}$ changes at most of about 15\%, relative to the
SSM ratio. Changes of this magnitude of the $\nu_B$ and $\nu_{Be}$ fluxes
are compatible with their theoretical uncertainties (about 10\% for the
$\nu_{Be}$ and about 30\% for the $\nu_B$ flux). Therefore, modifying
the temperature profile does not help much.

The {\em ${}^3$He profile} affects the neutrino fluxes much more strongly,
because the ${}^3$He abundance $Y_3(r)$ is not constrained
by helioseismology and is allowed to be considerably different from the SSM 
shape. One can consider in this case larger values of  $|\delta|$.

In Table~\ref{table2} we report the dependence of $I_{Be}$ and $I_B$ on
the ${}^3$He profile. Note that increasing the abundance in the center
relative to the outer part of the NPZ ($\delta < 0$) has the effect of
decreasing the ratio of the Be to B neutrino  flux. It is possible to
suppress this ratio by a factor almost 0.6 ($\delta=-4$) at the price,
however, of increasing (decreasing) the inner (outer) part of the NPZ
by a factor about 8 (1/4). Such a change of profile (a factor of thirty over
less than $0.1 R_{\odot}$) is really dramatic in comparison with the SSM
as one can see in Fig.~\ref{fig2}.
  
In the following we analyse why even large deformation of the ${}^3$He
profile can only partially improve the comparison with the
experiments. A deformation of the $^3$He radial profile that increases
the $^3$He abundance for $r < 0.0535 R_{\odot}$ and decreases it for
$r > 0.0535 R_{\odot}$ boosts $\Phi_{B}$ relative to $\Phi_{Be}$, since
the boron neutrinos are produces at smaller radii than beryllium ones.
However, such a deformation has the effect of moving both production
regions to smaller radii with two consequences: both fluxes increase
(even if by different factors) and the difference between the two
production regions shrinks. On the one hand, the more we deform the profile
the more is difficult to keep boosting $\Phi_{B}$ relative to $\Phi_{Be}$:
this explains why even the large deformation we have considered ($\delta=-4$
in our parameterization, which corresponds to an increase of $^3$He in
the inner part of the NPZ relative to the outer part by more than one order
of magnitude) can only improve the ratio by a factor
0.64 (see Table~\ref{table2}). On the other hand, even if a strong
deformation results in the required small $\nu_{Be}/\nu_B$ ratio,
it would be still necessary to strongly reduce both fluxes. This could
be achieve either by lowering the temperature or by lowering the overall
$^3$He abundance (the normalization factor $Y_{3,NPZ}$). The possibility of
lowering the temperature is limited by helioseismology, and it does not
seem physically possible to strongly reduce the $^3$He abundance in the
solar core (less strongly towards the center so that the required
profile deformation is achieved), since any mixing should {\em increase} 
the $^3$He in the core.

In Fig.~\ref{fig3} the thick solid line shows the variation of neutrino fluxes
with temperature in the case of the strongest deformation of $^3$He profile 
($\delta=-4$). The thin solid lines bound the region that is
spanned when also $S_{17}$ and $S_{34}$ are varied within the limits 
of Eqs.~(\ref{limit}).
The best fit point is labeled 4
($\Phi_{Be}=1.72\times10^9$ cm$^{-2}$s$^{-1}$ and 
$\Phi_{B}=2.41\times 10^6$ cm$^{-2}$s$^{-1}$); it is obtained with
$S_{34} = 0.38$~keV~b, $S_{17} = 31$~eV~b and $T_{NPZ}/T{NPZ}^{SSM} = 0.95$
and has a $\chi^2 = 32$, or about $5\sigma$'s away from the area allowed 
by experimental data. The  trajectory $2\to 3\to 4$ shows a possible way
to reach the best-fit point 4. 

Including contributions of CNO neutrinos to the signals makes the agreement
with experimental data even worse.

We could not reach in our calculations the point of Cumming and Haxton (shown 
in Fig.~\ref{fig1} and Fig.~\ref{fig3} by an asterisk) because of the 
seismic constraints imposed in our calculations and because we do not
allow an overall strong reduction of $^3$He in the core, which cannot be
caused by mixing.

\section{Conclusions}
The solar neutrino experiments have the status of disappearance oscillation 
experiments. This statement is based on the impossibility of explaining
the observed deficit of neutrino fluxes with astrophysical processes not
included in SSM's and revised nuclear cross-sections. Helioseismic
data strongly constrain the possible non-SSM astrophysical processes.     

Recently the idea of $^3$He mixing in the solar core has been revived 
\cite{Sch,Gough,Haxton,CummHaxt96}. The abundance of $^3$He is not directly 
constrained by seismic data. In principle this fact opens a road to
possible revisions of the SSM predictions for the neutrino fluxes.
However, in any realistic model, $^3$He mixing is accompanied by
other phenomena, {\em e.g.}, hydrogen is also transported into the core;
because of these accompanying phenomena $^3$He mixing can be
constrained by seismic observations (see~\cite{BPBCD,BTCM98}).
We present a more general approach where constraints on
the neutrino fluxes are valid for any mechanism of $^3$He mixing.

We assume arbitrary $^3$He mixing. Some unspecified process brings
fresh $^3$He from the $^3$He-richer outer shells into the solar core 
where thermal equilibrium is quickly established, but not
nuclear equilibrium. Since this process could consist of short
mixing episodes, we do not impose the solar-luminosity constraint.
The density radial profile is taken from seismic data. Within the
accuracy needed for calculations of neutrino fluxes, the $X(r)$ and $Y(r)$ 
profiles are also provided by seismic data. Therefore, the only
solar parameters that are needed for model-independent calculations of
neutrino fluxes are the temperature $T$, the $^3$He abundance $Y_3$,
the astrophysical factors $S_{34}$ and $S_{17}$ (cross-sections), and
the radial profiles of $T(r)$ and $Y_3(r)$.
We allow independent and large variations of $T$, $S_{34}$ and 
$S_{17}$ as given by Eqs.~(\ref{temp}) and (\ref{limit}). 

Regarding the radial dependence of the temperature, $T(r)$, and of the
$^3$He abundance, $Y_3(r)$, we used two approaches.

The first one consists in using the SSM radial dependences. This choice
appears reasonable, because B and Be neutrinos are actually produced in a
narrow region and distortions of the radial profiles, if not extreme, 
should not change the fluxes much. Results of this first approach are
presented in Fig.~\ref{fig1}.

The thick solid line describes the evolution of the fluxes with temperature. 
The two thin solid lines confine the region allowed by arbitrary  
variations of $^3$He abundance and the temperature, accompanied by variations
of $S_{34}$ and $S_{17}$ within the ranges given by Eqs.~(\ref{limit}).
The cross labeled 8 shows the best fit, which has a very large $\chi^2=35$.  

In the second approach we also allow changes in the shapes of the radial 
profiles $T(r)$ and $Y_3(r)$. We find
that the distortion of the $^3$He profile has a much stronger effect on the
neutrino fluxes and it is the only important one in our analysis. To improve
the agreement with experimental data, the $^3$He abundance should,
contrary to the SSM case, increase towards the center, as illustrated
by the dashed line in Fig.~\ref{fig2}. We are aware of no physical mechanism
in solar models that could produce such a dependence. We just take 
an unconvential and 
very strong radial dependence such as the one shown in Fig.~\ref{fig2}
as an {\em ad hoc} assumption.

With this extremely deformed $^3$He-density profile we repeat the exercise
of varying the $^3$He abundance and temperature in the neutrino production
zone,
with simultaneous variations of $S_{34}$ and $S_{17}$. In Fig.~\ref{fig3}
the allowed region is confined by the two thin solid curves.
The best fit to observational data is given by point 4,
and it still corresponds to a large $\chi^2=32$, {\em i.e.} more than
$5\sigma$ away from the experimentally allowed region.    

In all our calculations we have conservatively assumed a vanishing CNO
neutrino flux. Inclusion of any amount of CNO neutrinos makes the
disagreement with the experimental data worse.

We conclude that practically arbitrary $^3$He mixing, which also includes
physically unjustified distortions of the $^3$He radial profile, and
independent variations of temperature and cross-sections cannot explain
the observed solar neutrino fluxes.  

One might ask, however, whether and to what extent the considered 
uncertainties, especially in $^3$He mixing, can reduce the discrepancy
between the SSM solar-neutrino fluxes and observations. The most
important source of uncertainty is the $p+Be$ cross section ($S_{17}$), 
which affects
only the predicted flux of B neutrinos. Taking $S_{17}$ 40\% lower than 
presently used and the core temperature $T_c$ 1.4\% lower (as maximally
allowed~\cite{Ri} by seismic observations) one arrives at the minimum
B-neutrino flux $3\cdot 10^{6}$~cm$^{-2}$s$^{-1}$~\cite{Be}, {\em i.e.}
7.4$\sigma$ higher than the measured one. It is more difficult to estimate
a reasonable effect of $^3$He mixing (if this process exists at all).
It is constrained by accompaning processes, such as hydrogen mixing,
and a realistic model is needed for such calculations. The method used 
in this work can give only an upper limit of the influence of $^3$He
mixing on neutrino fluxes. The {\em ad hoc} assumptions used in this
analysis are maximally favorable for this influence, though rather
unrealistic.

\begin{figure}
  \caption[bbeex1]{
{\em Neutrino fluxes allowed by arbitrary $^3$He mixing accompanied
by independent variations of temperature, $S_{34}$ and $S_{17}$
(the $^3$He and temperature radial profiles are those of the SSM's). }
The horizontal (vertical) axis shows the beryllium (boron) flux; fluxes
are measured in units of the reference SSM of Bahcall and
Pinsonneault BP95~\cite{BP95} on the bottom (left) scale, and in
cm${}^{-2}$s${}^{-1}$ on the top (right) scale. The region 
allowed by the SSM's (BP95 SSM 90\% confidence region) is shown by the
dotted ellipse; the diamond shows the BBP98 SSM~\cite{BBP98} prediction with
the relative $1\sigma$ errors and the square shows the DS96~\cite{DS96}
SSM. The solid ellipse confines the region
allowed at $3\sigma$ by the experimental data (if there is any contribution
to the signals from CNO neutrinos this region becomes smaller).
A representative set of nonstandard solar model calculations are shown
by small ''x'' symbols and is taken from the review by Hata and
Langacker~\cite{HataLan95}, while the asterisk indicates the ${}^3$He mixing
assumption by Cunning and Haxton~\cite{CummHaxt96}.
The thick solid curve ``temp.'' gives the
evolution of the predictions of the BBP98~\cite{BBP98} model with
variations of temperature according to Eqs.~(\ref{parused}).
In particular points 1 and 4 correspond to temperatures lower than
$T_{SSM}$ by factors 0.98 and 0.95, respectively.
When $S_{17}$ diminishes, points move vertically down; when $S_{34}$
diminishes, they move diagonally towards the origin.
Three trajectories ($1\to 2\to 3$, $4\to 5$ and $4\to 6\to 7$) are shown for
illustration. The two thin solid curves bound the region allowed by
arbitrary changes of the $^3$He density and of the temperature, accompanied
by variations of $S_{34}$ and $S_{17}$ within the limits given by
Eqs.~(\ref{limit}). The best fit to the experimental data is labeled ``8''
and has the (too large) $\chi^2=35$.
  \label{fig1}
               }
\end{figure}
\begin{figure}
  \caption[he3]{
{\em Profiles of the $^3$He abundance normalized to 1 at
$R_{NPZ}=0.0535R_{\odot}$.}
The solid curve shows the profile in the SSM, while the dashed one shows
the profile corresponding to the largest deformation considered in this paper,
{\em i.e.}, $\delta= -4$ in Eq.~(\ref{profvar}) and in Table~\ref{table2};
the dotted curve is an example of different parameterization as discussed
in the text. 
\label{fig2}
               }
\end{figure}

\begin{figure}
  \caption[bbex3]{
{\em Same as Fig.~\ref{fig1}, but with a modified radial profile of 
the $^3$He abundance.}
This case corresponds to the maximal distortion ($\delta=-4$) of the 
SSM profile considered in this work (see Table~\ref{table2} and 
Fig.~\ref{fig2}). The temperature trajectory (towards lower temperatures) 
is shown by the thick solid curve ``temp.''. The points 1 and 2 correspond to 
temperatures lower than $T_{SSM}$ by factors 0.98 and 0.95, respectively.
The trajectory $2\to 3\to 4$ is shown for illustration: along
the track $2\to 3$ $S_{34}$ decreases by a factor 0.38/0.53, while
along the track $3\to 4$ $S_{17}$ increases by a factor 31/19.
The two thin solid curves bound the allowed region. For comparison
the region allowed
when keeping the SSM radial profile (see Fig.~\ref{fig1}) is bound by
the two dotted lines (thin solid lines in Fig.~\ref{fig1}).
The best fit is given by point 4 ($\chi^2= 32$), more than
$5\sigma$ away from the experimentally allowed region. 
  \label{fig3}
               }
\end{figure} 
\begin{table}
\caption[taa]{Dependence of the integrals $I_{Be}$ and $I_{B}$,
Eqs.~(\ref{intprobe}) and (\ref{intprob}),
on the temperature profile, parameterized according to
Eq.~(\ref{profvar}). The temperature profile  is deformed  keeping
$T(R_{NPZ})$ fixed. The first column shows the parameter
$\delta$ used in Eq.~(\ref{profvar}). The second and third columns give
the corresponding change of the integrals $I_{Be}$ and $I_{B}$, while
the fourth column shows the change of their ratio. The last two columns
illustrate the tilting of the profile relative to the SSM by reporting
$T/T^{SSM}$ at $ r=0.01 R_{\odot}$ and $r=0.1 R_{\odot}$.
\label{table1}
             }
\begin{tabular}{cccccc}
$\delta$ & 
$I_{Be}/I_{Be}^{SSM}$ & 
$I_{B}/I_{B}^{SSM}$ &
$(I_{Be}/I_{B})/(I_{Be}/I_{B})_{SSM}$ & 
$ T/T^{SSM} (0.01 R_{\odot}) $ &
$ T/T^{SSM} (0.1  R_{\odot}) $ \\
\tableline
-0.057 & 0.95 & 1.21 & 0.79 &  1.03& 0.98 \\
-0.044 & 0.96 & 1.14 & 0.84 &  1.02 & 0.98 \\
 0     & 1.0  & 1.0  & 1.0  &  1.0  & 1.0 \\
 0.044 & 1.08 & 0.94 & 1.15 &  0.98 & 1.02 \\
 0.057 & 1.11 & 0.94 & 1.18 &  0.97 & 1.02\\
\end{tabular}
\end{table}
\begin{table}
\caption[tbb]{
Dependence of the integrals $I_{Be}$ and $I_{B}$,
Eqs.~(\ref{intprobe}) and (\ref{intprob}),
on the profile of the ${}^3$He abundance. The information is analogous to the
one in Table~\ref{table1}.
\label{table2}
             }
\begin{tabular}{cccccc}
$\delta$ & $I_{Be}/I_{Be}^{SSM}$ & $I_{B}/I_{B}^{SSM}$ &
       $(I_{Be}/I_{B})/(I_{Be}/I_{B})_{SSM}$ & 
          $ y_3/y^{SSM}_3 (0.01 R_{\odot}) $ &
          $ y_3/y^{SSM}_3 (0.1 R_{\odot} )$ \\
\tableline
-4 & 1.20 & 1.87 & 0.64 &  8.06 & 0.24 \\
-3 & 1.05 & 1.51 & 0.70 &  4.78 & 0.34 \\
-2 & 0.98 & 1.27 & 0.77 &  2.84 & 0.49 \\
-1 & 0.96 & 1.10 & 0.87 &  1.69 & 0.70 \\
 0 & 1.0 & 1.0 & 1.0    &  1.0 & 1.0 \\
 1 & 1.11 & 0.95 & 1.17 &  0.59 & 1.43 \\
 2 & 1.32 & 0.93 & 1.41 &  0.35 & 2.06 \\
 3 & 1.66 & 0.96 & 1.72 &  0.21 & 2.95 \\
 4 & 2.23 & 1.03 & 2.16 &  0.12 & 4.24 \\
\end{tabular}
\end{table}
\begin{figure}[c]
\psfig{figure=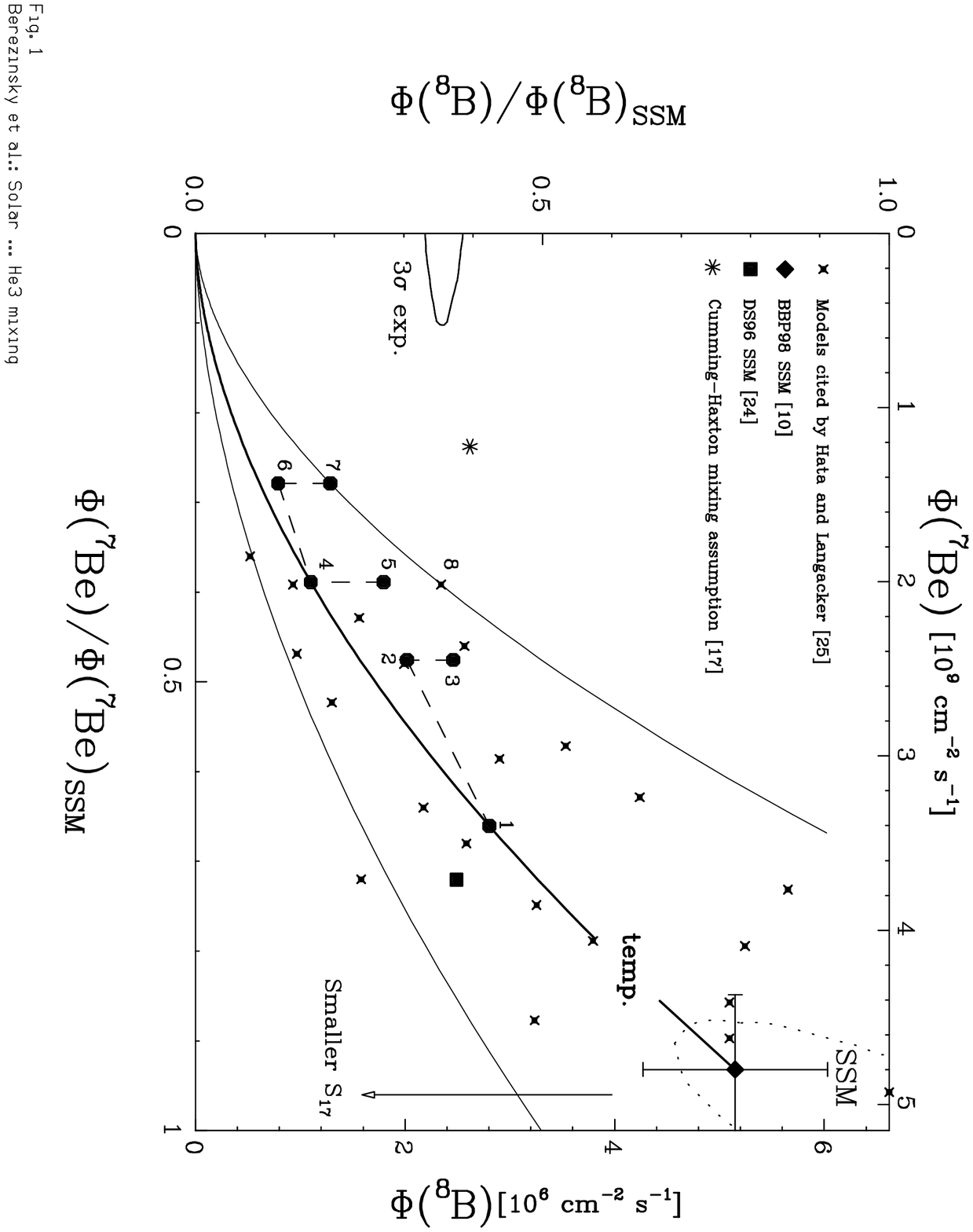,bbllx=36pt,bblly=36pt,bburx=576pt,bbury=756pt,%
height=22.5cm}
\end{figure}

\begin{figure}[c]
\psfig{figure=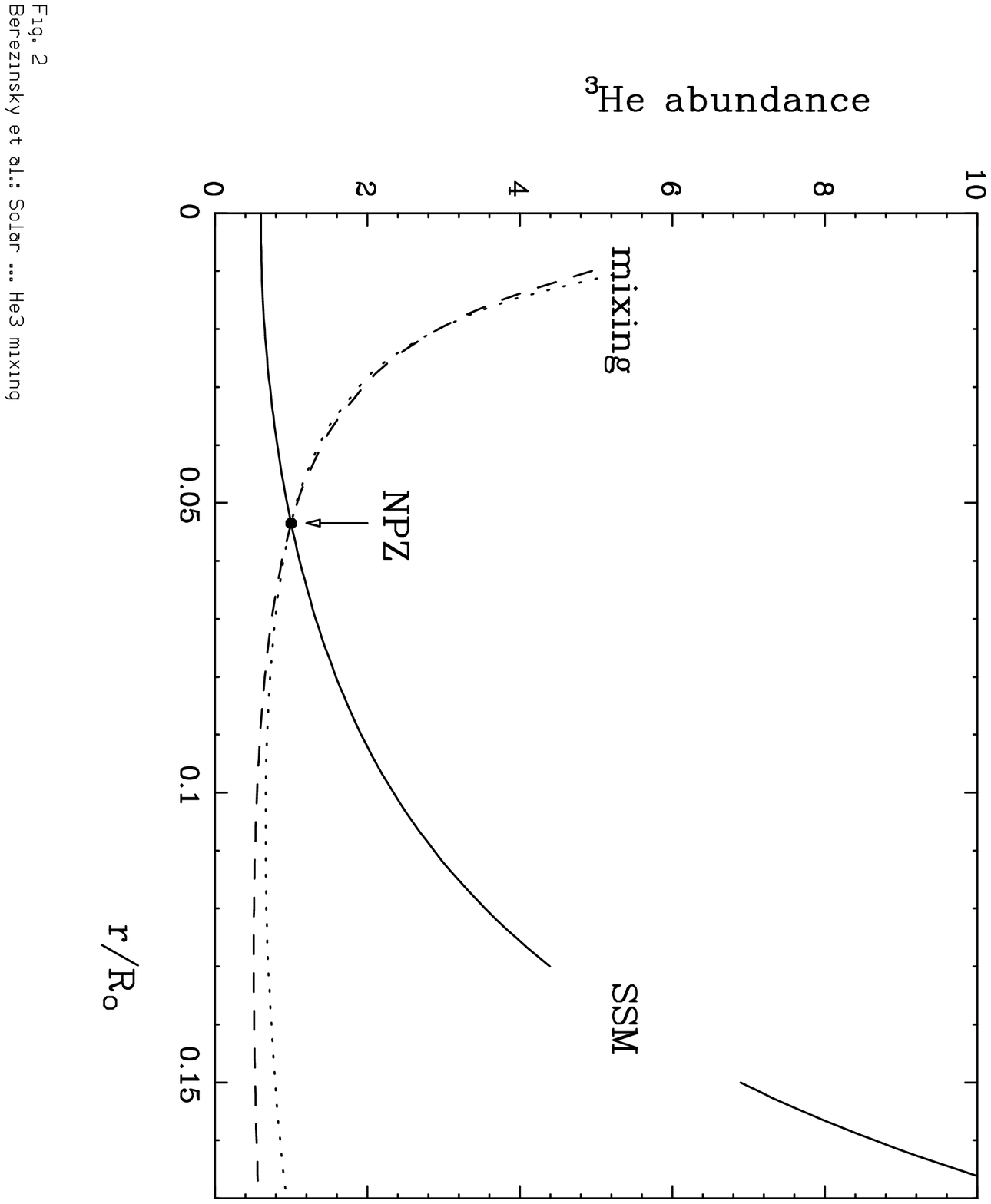,bbllx=36pt,bblly=36pt,bburx=576pt,bbury=756pt,%
height=22.5cm}
\end{figure}

\begin{figure}[c]
\psfig{figure=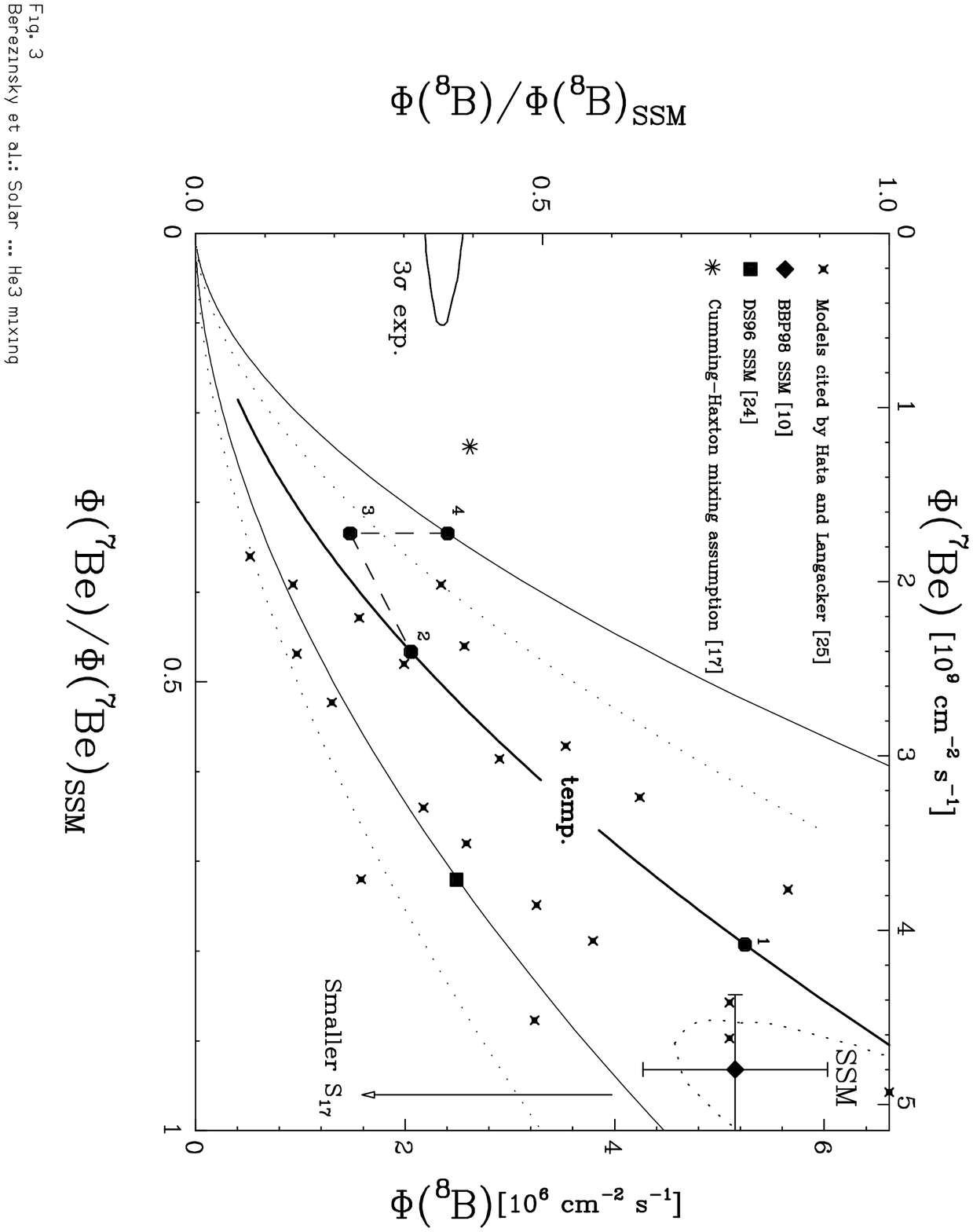,bbllx=36pt,bblly=36pt,bburx=576pt,bbury=756pt,%
height=22.5cm}
\end{figure}
\end{document}